\documentclass{PNASTWO}
\contributor{Submitted to Proceedings
  of the National Academy of Sciences of the United States of America}
\url{www.pnas.org/cgi/doi/}
\copyrightyear{ }
\issuedate{Issue Date}
\volume{Volume}
\issuenumber{Issue Number}

\begin{document}
\title{Vanishing of configurational entropy may not imply an ideal glass transition in randomly pinned liquids.}

\author{Saurish Chakrabarty\affil{1}
  {Centre for Condensed Matter Theory, Department of Physics,
    Indian Institute of Science, Bangalore, 560012, India},Smarajit Karmakar\affil{2}{TIFR Center for Interdisciplinary Science, Narsingi, Hyderabad 500075, India},\and
  Chandan Dasgupta\affil{1}{}\affil{3}{Jawaharlal Nehru Centre for Advanced 
    Scientific Research, Bangalore 560064, India.}}

\maketitle
\begin{article}
  Ref.~\cite{okim2014} presents numerical results for the configurational entropy density, $s_c$,  of a  model glass-forming liquid in the presence of random pinning.  The location of a ``phase boundary'' in the pin density ($c$) - temperature ($T$) plane, that separates an ``ideal glass'' phase from the supercooled liquid phase, is obtained by finding the points at which $s_c(T,c) \to 0$.  According to the theoretical arguments in Ref.~\cite{cammarotaPinning}, an ideal glass transition at which the $\alpha$-relaxation time $\tau_\alpha$ diverges takes place when $s_c$ goes to zero. 

\begin{figure*}[h]
    \begin{center}
      \vspace{-.4in}
      \includegraphics[height=0.52\columnwidth]{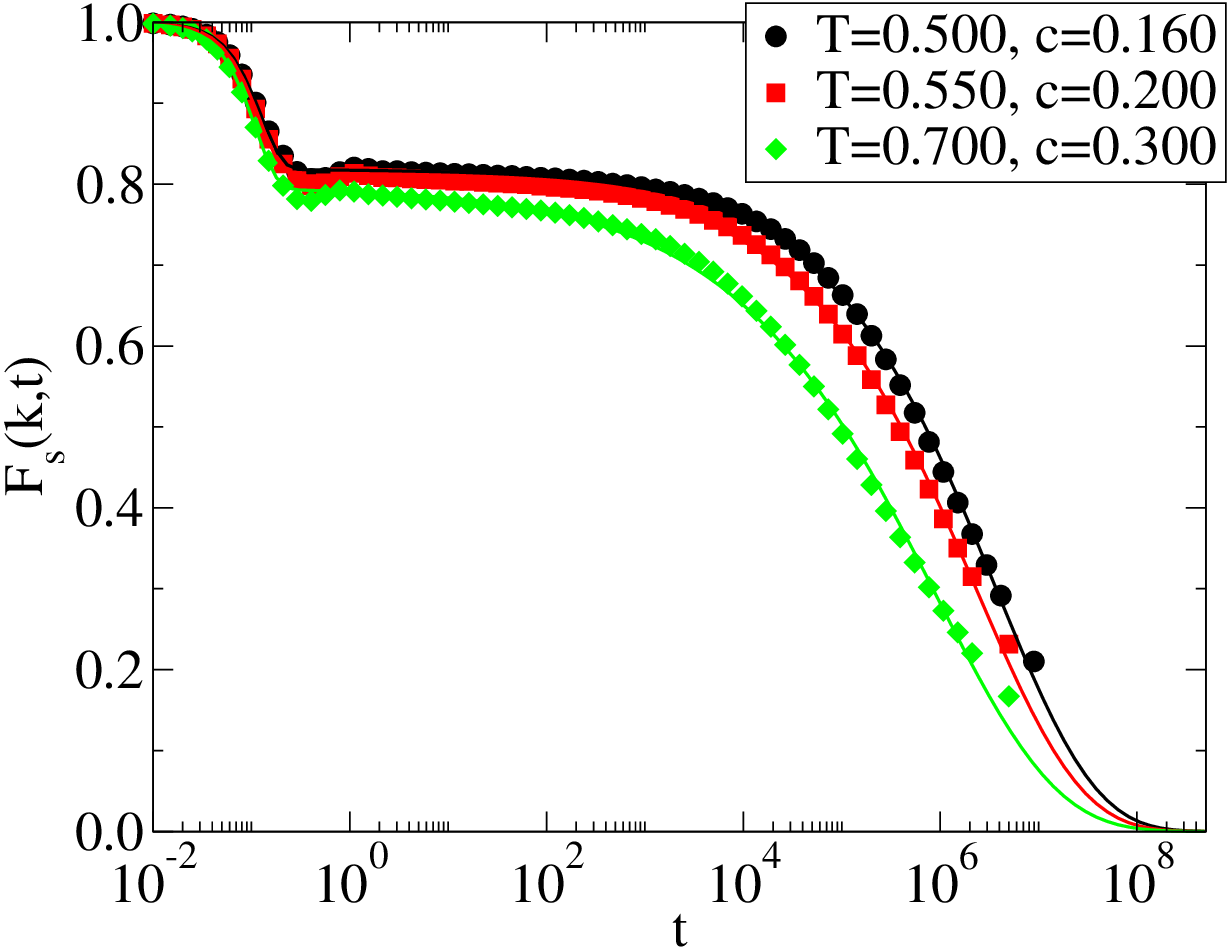}
      \includegraphics[height=0.52\columnwidth]{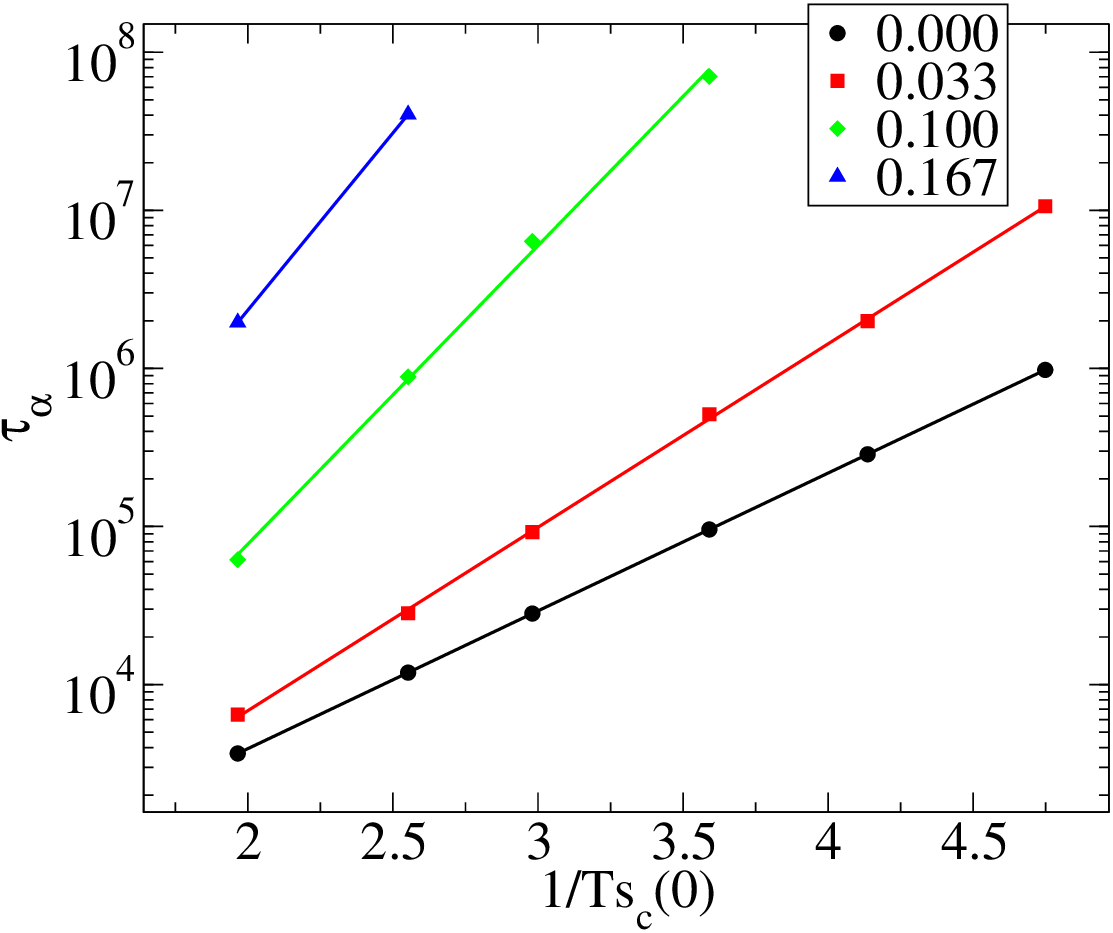} ~
      \includegraphics[height=0.52\columnwidth]{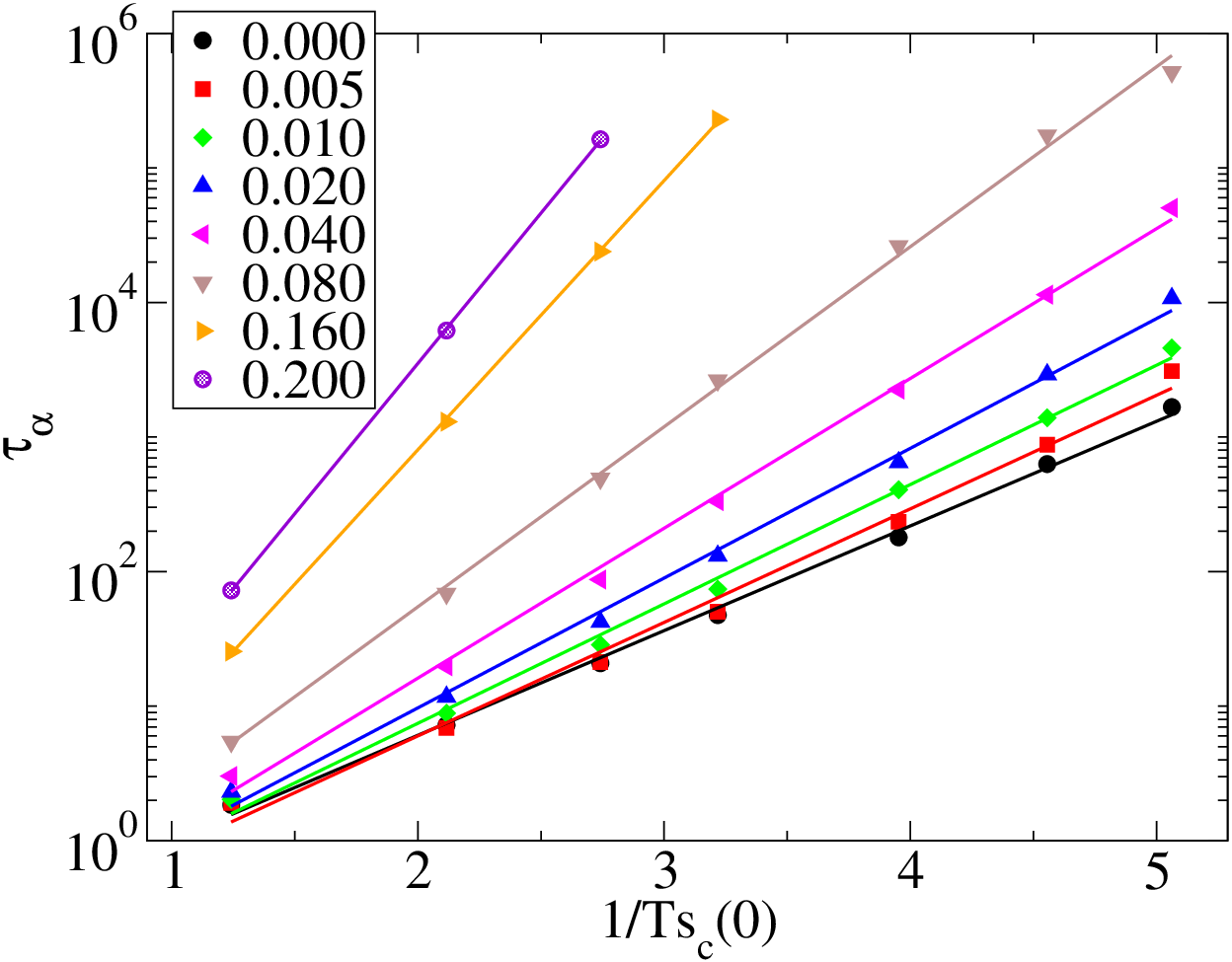}
      \caption{Left Panel: Plots of the time dependence of the self intermediate scattering
        function, $F_s(k,t) = (1/N_m)[<\sum_{i=1}^{N_m}
        e^{\imath\vec{k}\cdot\left(\vec{r}_i(t)-\vec{r}_i(0)\right)}>]$, 
        where $k$ is the wavenumber at the first peak of
        the static structure factor, $\langle\ldots\rangle$ implies an average over thermal history, 
        $[\ldots]$ represents an average over different realization of the pinned 
        particles, and $N_m$ is the number of unpinned mobile particles.
        Results are shown for three state points at which
        $s_c(T,c) \simeq 0$ according to Ref.~\cite{okim2014}. 
        The $\alpha$-relaxation time $\tau_{\alpha}$ is calculated
        using a fit to the form
        $F_s(k,t)=A\exp[-(t/\tau_1)^2/2]+(1-A)\exp[-(t/\tau_\alpha)^\beta]$. The fits are shown by solid lines.
        The relaxation times for these state points are: 
        $\tau_\alpha(T=0.50,c=0.16)\simeq 3.7\times10^6$,
        $\tau_\alpha(T=0.55,c=0.20)\simeq 2.4\times10^6$ 
        and $\tau_\alpha(T=0.70,c=0.30)\simeq 9.0\times10^5$.
        Middle Panel: $\ln[\tau_{\alpha}(T,c)]$ versus $1/[Ts_c(T,0)]$ for data
        extracted form Ref.~\cite{okim2014}. Right panel: $\ln[\tau_{\alpha}(T,c)]$ 
        versus $1/[Ts_c(T,0)]$ for
        the data in Ref.~\cite{ckdPinning2014}. 
        \label{fsktSc0}}
      \vspace{-.2in}
    \end{center}
  \end{figure*}

  We have studied the dynamics of the same system using molecular dynamics simulations. In Fig.\ref{fsktSc0} (left panel), we show the time-dependence of the self intermediate scattering function, $F_s(k,t)$,  calculated at three state points in the $(c-T)$ plane where $s_c(T,c) \simeq 0$ according to Ref.~\cite{okim2014}. It is clear from the plots that the relaxation time is finite [$\tau_\alpha \sim \mathcal{O}(10^6)]$ at these state points. Similar conclusions have been obtained in Ref.~\cite{ckdPinning2014} where an overlap function was used to calculate $\tau_\alpha$ at these state points.
  
  If the numerical results for $s_c(T,c)$ reported in Ref.~\cite{okim2014} are correct, then our explicit demonstration of the fact that $\tau_\alpha$ does not diverge at state points where $s_c=0$ according to Ref.~\cite{okim2014} would have fundamental implications for theories of the glass transition. The well-known Random First Order Transition (RFOT) description of the glass transition is based on the premise that the vanishing of $s_c$ causes a divergence of $\tau_\alpha$. The prediction~\cite{cammarotaPinning} of the existence of a line of ideal glass transitions in the $(c-T)$ plane for randomly pinned liquids was based on the RFOT description. Our results for $\tau_\alpha$ would imply that the RFOT description is {\it not valid} for pinned liquids. Since a divergence of $\tau_\alpha$ is the defining feature of the glass transition, the entropy-vanishing ``transition'' found in Ref.~\cite{okim2014} at which $\tau_\alpha$ does not diverge should not be called a glass transition.
  
  If, on the other hand, we disregard the results for $s_c(T,c)$ reported in Ref.~\cite{okim2014}, then all available data for the dynamics of this system~\cite{ckdPinning2014,jcppaper} can be understood from a description that is consistent with RFOT and the requirement that the presence of pinning must decrease $s_c$. This description~\cite{ckdPinning2014} is based on the assumption that $s_c(T,c) = B(c) s_c(T,0)$ for small values of $c$, where $B(c)$ is a smooth function that decreases with increasing $c$, with $B(0)=1$. This assumption, when combined with the Adam-Gibbs relation, predicts that the logarithm of $\tau_\alpha(T,c)$ should be a linear function of $1/[T s_c(T,0)]$ with a coefficient that increases with $c$. The data for $\tau_\alpha(T,c)$ in Ref.~\cite{ckdPinning2014} are consistent with this prediction (see Fig.\ref{fsktSc0}, right panel). We have verified that the data for $\tau_\alpha (T,c)$ and $s_c(T,0)$ in Ref.~\cite{okim2014} are also consistent with this prediction (see Fig.\ref{fsktSc0}, middle panel). This observation provides a way of reconciling the behavior of $\tau_\alpha(T,c)$ with RFOT, but it also implies that the data for $s_c(T,c)$ reported in Ref. \cite{okim2014} are not quantitatively accurate.


\begin{thebibliography}{1}

  \bibitem{okim2014}
    Ozawa, M, Kob, W, Ikeda, A,  \& Miyazaki, K.
    \newblock (2015) Equilibrium phase diagram of a randomly pinned glass-former.
    \newblock {\em Proc. Nat'l Acad. Sci. USA} p. 201500730.

  \bibitem{cammarotaPinning}
    Cammarota, C \& Biroli, G.
    \newblock (2012) Ideal glass transitions by random pinning.
    \newblock {\em Proc. Nat'l Acad. Sci. USA} {\bf 109}, 8850--8855.

  \bibitem{ckdPinning2014}
    Chakrabarty, S, Karmakar, S,  \& Dasgupta, C.
    \newblock (2015) Dynamics of glass forming liquids with randomly pinned
    particles.
    \newblock {\em arXiv:1404.2701}, to be published in Scientific Reports.

  \bibitem{jcppaper}
    Yan-Wei Li, You-Liang Zhu \& Zhao-Yan Sun.
    \newblock (2015) Decoupling of relaxation and diffusion in random pinning
    glass-forming liquids.
    \newblock {\em J. Chem. Phys} {\bf 142}, 124507.



  \end{thebibliography}
  
\end{article}
\end{document}